\documentclass[preprint,showpacs,amsmath,amssymb,aip,jap,floatfix]{revtex4-1} 
\usepackage{graphicx}
\usepackage{dcolumn}
\usepackage{bm}
\usepackage{braket}
\usepackage{mathtools}
\usepackage[utf8]{inputenc}
\usepackage[T1]{fontenc}
\usepackage{mathptmx}
\usepackage{hyperref}
\usepackage{color}
\usepackage[table,xcdraw]{xcolor}

\begin{document}

\title{Electron-electron double resonance detected NMR spectroscopy using ensemble NV centers at 230 GHz and 8.3 Tesla}

\author{Benjamin Fortman}
\email{bfortman@usc.edu}
\affiliation{Department of Chemistry, University of Southern California, Los Angeles CA 90089, USA}

\author{Laura Mugica-Sanchez}
\affiliation{Department of Chemistry, University of Southern California, Los Angeles CA 90089, USA}

\author{Noah Tischler}
\affiliation{Department of Chemistry, University of Southern California, Los Angeles CA 90089, USA}

\author{Cooper Selco}
\affiliation{Department of Physics \& Astronomy, University of Southern California, Los Angeles CA 90089, USA}

\author{Yuxiao Hang}
\affiliation{Department of Physics \& Astronomy, University of Southern California, Los Angeles CA 90089, USA}

\author{Karoly Holczer}
\affiliation{Department of Physics \& Astronomy, University of California, Los Angeles CA 90095, USA}

\author{Susumu Takahashi}
\affiliation{Department of Chemistry, University of Southern California, Los Angeles CA 90089, USA}
\affiliation{Department of Physics \& Astronomy, University of Southern California, Los Angeles CA 90089, USA}

\date{\today}

\begin{abstract}
The nitrogen-vacancy (NV) center has enabled widespread study of nanoscale nuclear magnetic resonance (NMR) spectroscopy at low magnetic fields.
NMR spectroscopy at high magnetic fields significantly improves the technique's spectral resolution, enabling clear identification of closely related chemical species. 
However, NV-detected NMR is typically performed using AC sensing through electron spin echo envelope modulation (ESEEM), a hyperfine spectroscopic technique that is not feasible at high magnetic fields.
Within this paper, we have explored an NV-detected NMR technique for applications of high field NMR.
We have demonstrated optically detected magnetic resonance (ODMR) with the NV Larmor frequency of 230 GHz at 8.3 Tesla, corresponding to a proton NMR frequency of 350 MHz.
We also demonstrated the first measurement of electron-electron double resonance detected NMR (EDNMR) using the NV center and successfully detected $^{13}C$ nuclear bath spins.
The described technique is limited by the longitudinal relaxation time ($T_1$), not the transverse relaxation time ($T_2$). 
Future applications of the method to perform nanoscale NMR of external spins at 8.3 T and even higher magnetic fields are also discussed.
\end{abstract}

\maketitle

\section{Introduction}
Magnetic resonance techniques, such as nuclear magnetic resonance (NMR) and electron spin resonance (ESR), provide exquisite information about local chemical environments.
NMR spectroscopy is routinely used in chemical synthesis for structural analysis of small molecules.
The nitrogen-vacancy (NV) center is an excellent candidate for nanoscale magnetic resonance as its spin state can be optically initialized and readout, has demonstrated long coherence times, and is highly sensitive to external magnetic fields.~\cite{Gruber1997, Jelezko2004,Childress2006, Takahashi2008, Bauch2018, Wolf2015, Schmitt2017}
The NV center's unique properties have enabled both nanoscale NMR and ESR with sensitivity down to the level of a single spin.~\cite{Muller2014,Shi2018,Shi2013,Sushkov2014,Abeywardana2016}
NV-detected NMR is now widely used at low magnetic fields ($<0.1$ T), such as for NV depth estimation, liquid state NMR, two-dimensional NMR, hyperpolarized NMR, nanodiamond based NMR, and even for selective spin manipulation in a 10-qubit quantum register.~\cite{Pham2016,Kehayias2017,Smits2019,Holzgrafe2020,Bradley2019}

NMR at high magnetic fields greatly increases the spectral resolution and improves sensitivity. 
The increase in field strength increases the frequency difference between closely related chemical species and enables resolution of small chemical shifts. 
High field NMR offers new insights into molecules with many similar nuclei, low gyromagnetic ratios, and low natural abundance, such as for $^{17}O$ NMR in pharmaceutical compounds and biomacromolecules.~\cite{Kong2013,Keeler2019}
Commercial NMR magnets operating at 28.2 T (proton Larmor frequency of 1.2 GHz) have recently become available, with hybrid magnets at fields of 35.2 T (corresponding to 1.5 GHz proton NMR) being available in user facilities.~\cite{Luchinat2021, Gan2017}
Implementation of NV-detected NMR at a high magnetic field is highly desirable.
However, there have only been a handful of studies on NV-based sensing at high magnetic fields due to technological challenges involved with combining a NV ODMR system with a high magnetic field ESR system.~\cite{Stepanov2015,Fortman2020,Aslam2015,Aslam2017}

In this paper, we discuss the implementation of NV-detected NMR at a high magnetic field.
NV-detected NMR can be achieved by hyperfine spectroscopic techniques.
There are three primary pulsed ESR hyperfine spectroscopic techniques: electron spin echo envelope modulation (ESEEM), electron-nuclear double resonance (ENDOR), and electron-electron double resonance detected NMR (EDNMR).~\cite{schweiger2001principles, Mims1965,Schosseler1994} 
Most NV-detected NMR spectroscopy performed at a low magnetic field is based on ESEEM where the hyperfine coupling between the NV center and nuclear spins mixes the spin-state and results in periodic revivals of the echo intensity.~\cite{Childress2006}
This technique functions very efficiently only when the energies of the hyperfine coupling and the nuclear Larmor frequency are comparable.
Therefore, it works well at a low magnetic field, but becomes unfeasible at a high magnetic field.~\cite{schweiger2001principles}
In ENDOR and EDNMR techniques, the population difference of an ESR transition is monitored via a detection scheme while either pulsed RF (ENDOR) or off resonance MW (EDNMR) radiation is applied to drive polarization transfer. 
Nuclear identification and determination of hyperfine coupling is performed based on the frequency of polarization transfer.
EDNMR uses a high turning angle (HTA) pulse to drive population transfer on forbidden transitions. 
At higher fields, the Zeeman interaction more completely dominates over the hyperfine interaction, reducing state mixing and consequently, transition probability.
Therefore, higher magnetic fields require stronger or longer HTA pulses to induce polarization transfer.
Both EDNMR and ENDOR are promising for NV-based sensing, as they are applicable to single and ensemble NV systems and are limited by the longitudinal relaxation time, $T_1$, instead of the transverse relaxation time, $T_2$. 
The $T_1$ relaxation time for NV ensembles has been shown to extend dramatically (up to minutes) at low temperature.~\cite{Takahashi2008,Jarmola2012}
More recently, EDNMR has emerged as a promising technique due to its higher sensitivity and resiliency against RF related artifacts.~\cite{Cox2017} 
EDNMR has an additional advantage over ENDOR in that it does not require an additional RF power amplifier or tuned RF circuit and can thus be readily implemented over a large frequency range for the detection of nuclei with a wide range of gyromagnetic ratios. 

Within this work, we demonstrate optically detected magnetic resonance (ODMR) on the NV center at the highest field and frequency to date, 8.3 T, corresponding to the NV's Larmor frequency of 230 GHz (proton Larmor frequency of 350 MHz).
We successfully implement EDNMR using ensemble NV centers and detect $^{13}C$ nuclear bath spins in the diamond crystal.
Since the EDNMR technique is limited by $T_1$, not $T_2$, NV-detected NMR based on EDNMR can take advantage of the NV center's long $T_1$ to perform measurements with a long HTA pulse.
 With development of suitable pulse capabilities, the described NV-detected NMR technique will be advantageous for the development of NV-detected NMR at higher fields and frequencies where the microwave power is often limited.~\cite{Morley2008, Takahashi12, Fortman2020}

\section{Methods and Materials}
\begin{figure}
    \centering
    \includegraphics[width=3.3in]{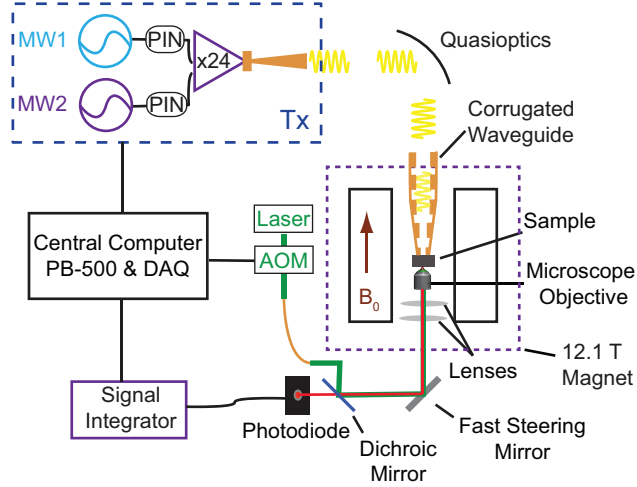}
    \caption{Overview of the experimental setup. 
    The transmission (Tx) setup consists of two independently controllable frequency sources (MW1 and MW2) that pass through PIN switches to a frequency multiplication chain. 
    High frequency MW excitation is propagated through quasioptics and a corrugated waveguide to the sample stage within a 12.1 Tesla variable field magnet. 
    Pulsed laser excitation is directed through an acousto-optic modulator (AOM) and an optical fiber to a system of lenses, a fast steering mirror, and the sample stage. 
    At the sample stage, a microscope objective directs laser intensity and collects sample fluorescence. 
    The fluorescence is redirected through a dichroic mirror to a photodiode where it is integrated using either gated boxcar integrators or a fast oscilloscope.
    The MW components, laser, and boxcar integrators are all controlled through a central computer equipped with a fast TTL logic board and digital to analog converter (DAQ).
    The magnetic field ($B_0$) is aligned with the optical axis.
    }
    \label{fig:Setup}
\end{figure}
A home-built, high field (HF) ODMR spectrometer operating in the band of 215-240 GHz was used.
An overview of the experimental setup is shown in Fig.~\ref{fig:Setup}.
The diamond sample was mounted at the center of a variable field 12.1 T superconducting magnet (Cryogenic Limited).
Microwave (MW) excitation was produced by a solid state source (Virginia Diodes) and directed through quasioptics to the sample stage.
The output power of both channels from the source was 115 mW at 230 GHz.
Laser excitation was produced from a solid-state single mode laser (Crystalaser) and directed through an acousto-optic modulator (Isomet), single mode fiber (Thorlabs), and microscope objective (Zeiss100X, NA=0.8) before reaching the sample stage. 
The excitation beam position was controlled using a fast steering mirror (Newport) and a system of lenses below the microscope objective.
Fluorescence (FL) collected at the objective was directed back through a dichroic mirror and fluorescence filters (Omega Optics) before being detected using a photodiode (Thorlabs 130A2).
The typical excitation spot size was a few $\mu$m$^2$.
Typical laser excitation of $\sim 4$ mW at the sample stage resulted in 1-2 $\mu$W of detected FL. 
The output of the photodiode was directed to a signal integrator. 
Integration was performed using either a pair of analog boxcar integrators (Stanford Research Systems SR250) or a fast digitizing oscilloscope (Tektronix MSO64B). The analog output of the boxcar integrators was digitized using a fast DAQ (National Instruments  PCIe-6321).
Gate timing was controlled using a gated TTL logic board (SpinCore Technologies PB-500). 
Additional details of the HF-ESR/ODMR spectrometer have been described previously.~\cite{Cho2014,Cho2015, Stepanov2015, Fortman2020}
For this study, two samples were used. Sample 1 was a 2.0 $\times$ 2.0 $\times$ 0.3 mm$^3$ size, (111)-cut high pressure, high temperature type Ib diamond from Sumitomo Electric Industries. Sample 2 was a hexagonal 4.4 $\times$ 3.9 $\times$ 0.5 mm$^3$ size, (111)-cut  high pressure high temperature type-Ib diamond obtained from Element Six.
Both diamonds had previously been subjected to high energy (4 MeV) electron beam irradiation and were exposed to a total fluence of $1.2 \times 10^{18}$ e$^-/$cm$^2$ followed by an annealing process at 1000 $^o$C. This treatment  produced a NV concentration greater than 1 ppm.~\cite{Fortman2020}

\section{Discussion}
We begin by performing pulsed ODMR on ensemble NV centers. 
\begin{figure}
    \centering
    \includegraphics[width=3.3in]{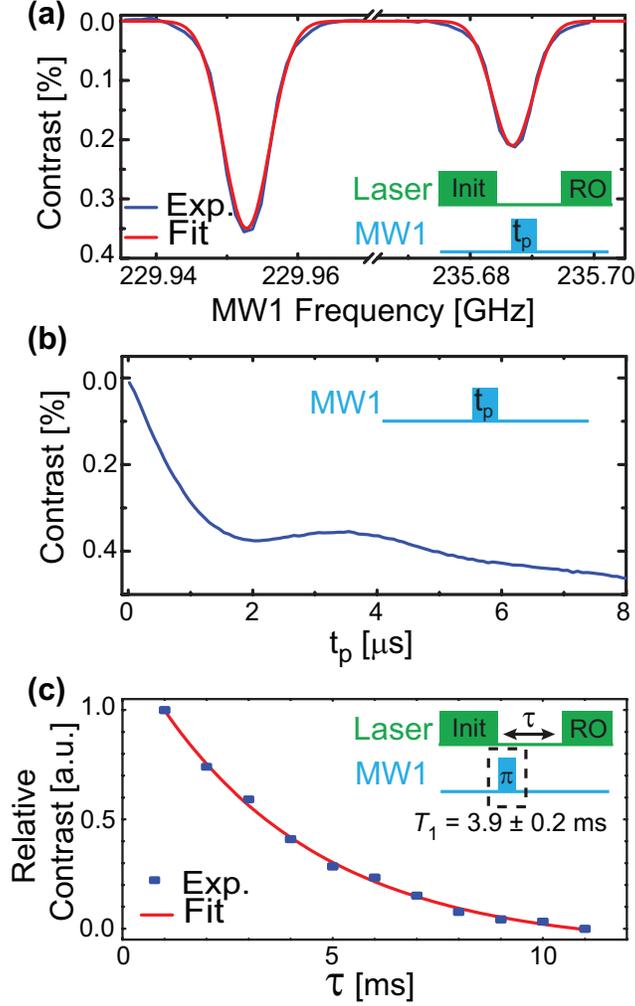}
    \caption{Ensemble ODMR at 230 GHz.
    (a) Pulsed ODMR data. For all ODMR measurements, laser pulses of $20$ $\mu$s and $15$ $\mu$s were used for initialization (Init) and readout (RO), respectively. 
    After initialization, a MW1 pulse ($t_p$) of $1.9$ $\mu$s was applied and varied in frequency. 
    Clear reductions in FL intensity were resolved at $229.953$ GHz and $235.687$ GHz, corresponding to the lower ($\ket{m_S=0} \leftrightarrow \ket{m_S=-1}$) and upper ($\ket{m_S=0} \leftrightarrow \ket{m_S=+1}$) transitions of the NV center.
    The magnetic field was found to be $8.306$ T with a polar angle of $1.50 \pm 0.02^\circ$.
    Fitting was performed using nonlinear least squares regression and the NV center Hamiltonian ($S = 1$, $D = 2870$ MHz, $g = 2.0028$).~\cite{Takahashi2008}
    (b) Measurement of Rabi oscillations. 
    The frequency of MW1 was set at the lower resonance and the pulse length was varied. 
    From the observed oscillations, a $\pi$ pulse length of $1.9$ $\mu$s was found. 
    (c) Measurement of $T_1$ relaxation. Measurements were performed with (Sig1) and without (Sig2) a $\pi$ pulse.
    The difference (Sig2-Sig1) was normalized and then fit to a single exponential decay.~\cite{Jarmola2012}
    Data was collected using (a) 10 scans, (b) 18 scans, and (c) Sig1 and Sig2 were measured sequentially with 5 scans each.
    }
    \label{fig:NVCharacterization}
\end{figure}
For pulsed ODMR, the relative FL intensity was monitored while a MW pulse was varied in frequency. 
As seen in Fig.~\ref{fig:NVCharacterization}(a), clear reductions in FL intensity were resolved at $229.953$ GHz and $235.687$ GHz, corresponding to the lower ($\ket{m_S=0} \leftrightarrow \ket{m_S=-1}$) and upper ($\ket{m_S=0} \leftrightarrow \ket{m_S=+1}$) transitions of a [111] oriented NV with a polar offset angle of $1.50 \pm 0.02$ degrees.
Next, Rabi oscillations were recorded by fixing the frequency of MW1 at $229.953$ GHz ($\ket{m_S=0} \leftrightarrow \ket{m_S=-1}$ transition) and varying the pulse length as seen in Fig.~\ref{fig:NVCharacterization}(b). 
From these measurements, damped oscillations and a $\pi$ pulse length of $1.9$ $\mu$s was observed.
Next, the NV ensemble's spin-lattice relaxation time, $T_1$, was recorded. 
For this measurement, the duration between the laser initialization and readout pulse ($\tau$) was varied (see Fig.~\ref{fig:NVCharacterization} (c). 
Two sequential measurements were performed by varying the spacing between initialization and readout with and without a MW $\pi$ pulse before normalization. A $T_1$ time of $3.9 \pm 0.2$ ms was found by fitting to a single exponential decay.  

\begin{figure*}
    \centering
    \includegraphics[width=6in]{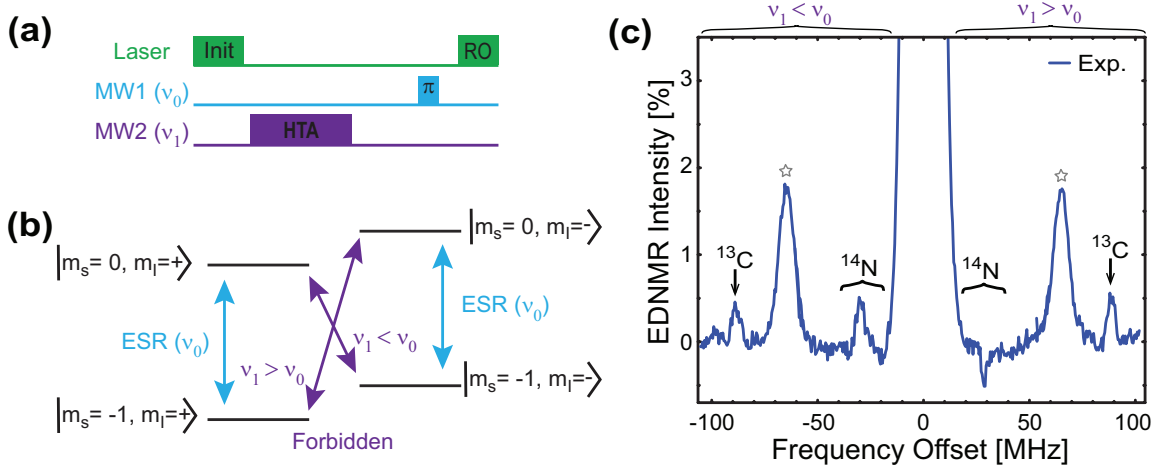}
    \caption{NV detected EDNMR at high field.
    (a) Pulse sequence used in the NV-detected EDNMR experiment.
    In the experiment, a HTA pulse was applied with MW2 at frequency $\nu_1$ before a  $\pi$ pulse was applied with MW1 at frequency $\nu_0$. 
    The frequency of $\nu_0$ was set to match the lower transition.
    The application of a $\pi$ pulse increases the sensitivity by isolating the FL of [111] oriented NV centers from non axial orientations.
    (b) Energy level diagram. 
    Nuclei coupled via weak hyperfine interaction are represented by $m_I = +$ and $m_I = -$. 
    During the experiment, the frequency of the HTA pulse is swept from below ($\nu_1<\nu_0$) to above ($\nu_1>\nu_0$) the central ESR resonance. 
    Population is transferred when the HTA pulse is in resonance with the difference between coupled states, resulting in an increase in the observed FL.
    Due to the length of the HTA pulse and state mixing induced by the hyperfine interaction, this occurs for both allowed and forbidden transitions.
    The intensity of the central blind spot is due to the allowed transitions. 
    (c) Experimental spectra.
    The data are shown with reference to the MW frequency offset ($\nu_1 -\nu_0$) and  normalized to the intensity of the central blind spot.
    In the present case, $\nu_0 = 229.9528$ GHz.
    A 500 $\mu$s HTA pulse and 1.9 $\mu$s $\pi$ pulse were used.
    The length of the HTA pulse was chosen to minimize the influence of $T_1$ relaxation after population transfer.
    EDNMR signals due to forbidden transitions involving $^{14}N$ and $^{13}C$ are indicated.
    Grey stars are used to indicate peaks due to allowed transitions from nearest neighbor $^{13}C$ lattice sites.
    For (c), data was collected using 20 scans over a period of 11 hours.
    }
    \label{fig:HFNVEDNMR}
\end{figure*}

Next we perform EDNMR using the NV center.
As shown in Fig.~\ref{fig:HFNVEDNMR}(a), EDNMR is a form of high field hyperfine spectroscopy that utilizes two microwave frequencies, MW1 ($\nu_0$) and MW2 ($\nu_1$).
EDNMR measurements vary the frequency ($\nu_1$) of a HTA MW2 pulse, while MW1 applies a detection pulse sequence, such as Hahn echo, at $\nu_0$ to measure the spin polarization of an ESR transition.~\cite{Schosseler1994} 
As the frequency of $\nu_1$ is swept, the frequency shifts on resonance with transitions below the central transition ($\nu_1<\nu_0$) due to weakly coupled hyperfine nuclei, as seen in Fig.~\ref{fig:HFNVEDNMR}(b).
These transitions are generally forbidden as they involve a flip of both the electron and nuclear spin ($\Delta m_S = 1, \Delta m_I = 1$). 
The forbidden transitions become weakly allowed with partial state mixing, leading to polarization transfer and a reduction in the ESR signal intensity.
This change is detected as an EDNMR signal.
Application of a long HTA pulse improves the likelihood of population transfer, but the total length of the HTA pulse must be short relative to $T_1$ in order to maximize the observable contrast.
As $\nu_1$ approaches the central allowed transition ($\Delta m_S = 1,\Delta m_I = 0$) there is significant population transfer leading to a highly intense change and the so-called "central blind spot".
Since the central blind spot highly distorts EDNMR signal, in practice the measurement is performed at a frequency range outside of the central blind spot.
After passing the central blind spot, $\nu_1$ then induces forbidden transitions from hyperfine coupled nuclei with a positive frequency offset ($\nu_1>\nu_0$) relative to the central transition.
For NV detected EDNMR, the spin population can be directly detected via optical spin state readout, eliminating the need for an echo detection sequence. 
EDNMR with the NV center has an advantage over conventional EDNMR, as optical initialization of the NV center ensures high spin polarization and improves EDNMR sensitivity. 
The usage of optical initialization shortens the measurement time by eliminating the need for long cycle delays between subsequent experiments (typically $\gg$ $T_1$).

 As shown in Fig.~\ref{fig:HFNVEDNMR}(a), we perform the experiment by applying an initialization laser pulse, MW2 HTA pulse at frequency $\nu_1$, MW1 $\pi$ pulse at frequency $\nu_0$, and laser readout pulse.
 During the experiment $\nu_1$ is varied while $\nu_0$ is fixed at the lower NV resonance.
When the HTA pulse drives a transition, the population of the $\ket{m_S = 0}$ spin state is reduced before the MW1 $\pi$ pulse transfers the population to the $\ket{ms=-1}$ state.
Therefore, when the HTA pulse is in resonance with a transition, an increase in the FL intensity is observed.
For the present experiment, a HTA pulse length of $500$ $\mu$s was chosen.
In principle, longer length pulses, up to $T_1$, can be applied.
Figure~\ref{fig:HFNVEDNMR}(c) shows the result of the experiment and we observe signals at $\pm88$, $-64$, $-30$, $+28$, and $+65$ MHz. 
The strong change in the FL intensity at $0$ MHz corresponds to the central blind spot.
The signals at $-64$ and $+65$ MHz give the hyperfine coupling constant of 129 MHz, consistent with nearest neighbor $^{13}C$ hyperfine interaction ($126$-$130$ MHz) splitting the allowed ESR transition.~\cite{Nizovtsev2010,Smeltzer2011,Jarmola2016}
The reduced intensity relative to the central transition corresponds to the low natural abundance of $^{13}C$ ($\sim$1.1\%) and low probability of nearest neighbor locality.

\begin{figure*}
    \centering
    \includegraphics[width=6.3in]{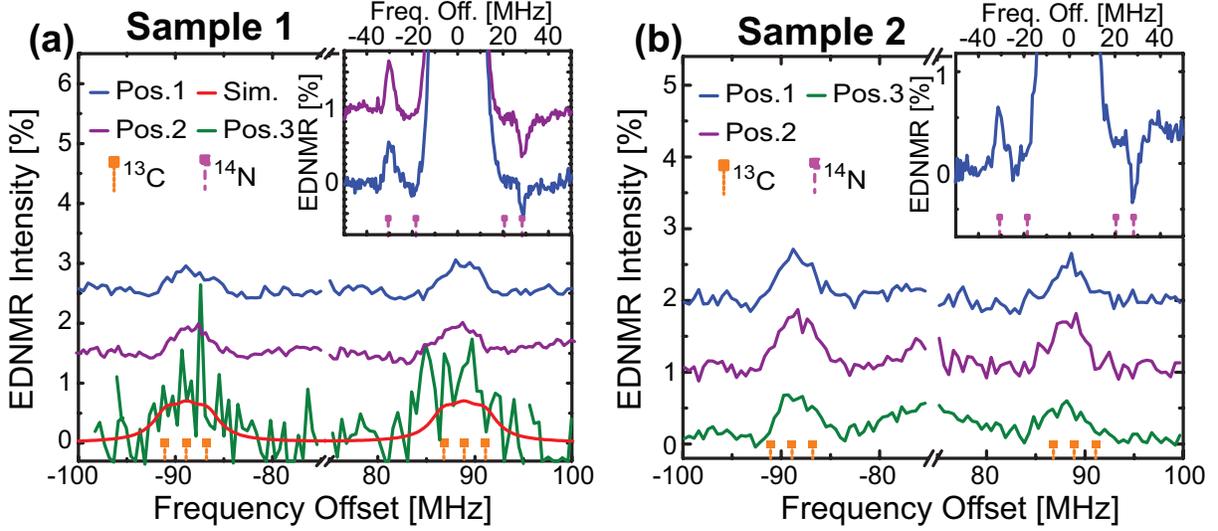}
    \caption{
    NV detected EDNMR at 8.3 Tesla
    (a) NV detected EDNMR from sample 1. 
    EDNMR detection of $^{13}C$ is shown in the main graph and EDNMR detection of $^{14}N$ is shown in the inset.
    Data is offset for clarity. The presented data is from three areas: positions 1 and 2 were spaced $\sim 50$ $\mu$m apart, position 3 was taken after removing and replacing the sample. The data for position 3 was integrated with boxcar integrators, all other measurements were integrated using the fast oscilloscope.
    For position 3, variations in the experimental setup resulted in slightly different parameters: the magnetic field was 8.298 T with a polar offset angle of $1.9 \pm 0.1^o$. Rabi oscillations showed a $\pi$ pulse length of $1.6$ $\mu$s. The change in magnetic field resulted in a small ($\sim 0.1$ MHz) shift in the transition frequencies.
    The stick spectrum shows double quantum transitions from Table \ref{tb:NVTransitions}.
    A simulation based upon $^{13}C$ coupling to the NV center is shown in red.
    The red line shows a simulation of $L(\omega;\Delta \omega,\omega_i) = A/\pi \sum_{\omega_i} {\Delta \omega}/({\Delta \omega^2+4(\omega - \omega_i)^2})$ where $A$ is an amplitude and the sum runs over the resonance positions ($\omega_i$).
    A nitrogen spin concentration of 70 ppm ($\Delta \omega = 3.2$ MHz) was used in agreement with sample properties.
    Nonlinear regression of $L(\omega;\Delta \omega,\omega_i)$ was used to determine $\Delta \omega$ from the experimental data (fits not shown). $\Delta \omega$ was measured to be $2.3 \pm 0.3 $, $2.9 \pm 0.4 $, and $3.7 \pm 0.8 $ MHz for positions 1, 2, and 3 respectively.
    (b) NV detected EDNMR from sample 2. 
    EDNMR detection of $^{13}C$ is shown in the main graph and EDNMR detection of $^{14}N$ is shown in the inset. Data is offset for clarity.
    The presented data is from three areas: position 1, 2 and 3 were spaced $\sim 50$ $\mu$m apart from each other. The stick spectrum shows the position of double quantum transitions.
    For sample 2, $\Delta \omega$ was measured to be $2.7 \pm 0.3 $, $2.9 \pm 0.3 $, and $2.5 \pm 0.3 $ MHz for positions 1, 2, and 3 respectively.
    }
    \label{fig:C13EDNMR}
\end{figure*}

Next we discuss signals at $\pm 88$ MHz.
In order to understand the signals we discuss the following Hamiltonian:
\begin{equation}
H_{NV} = \mu_B g_{NV} \Vec{B_0} \cdot \Vec{S} + D \Vec{S_z}^2 +H_{N} + H_{C},
    \label{eq:NVHamil}
\end{equation}
where $D = 2.87$ GHz, $g_{NV} =2.0028$, and $\Vec{S}$ is the electronic spin operator.~\cite{Takahashi2008} $H_{N}$ and $H_{C}$ represent the Hamiltonians of hyperfine coupled nitrogen in the NV center and surrounding $^{13}C$ bath spins.
The nuclear spin Hamiltonians may be written as:
\begin{subequations}
\begin{align}
H_{N} = -\gamma_{^{14}N} \Vec{B_0} \cdot \Vec{I_1} +\Vec{S} \cdot \Vec{A}_{^{14}N} \cdot \Vec{I_1} + P I_{1z}^2 \label{H14N},\\
H_{C} = -\gamma_{^{13}C} \Vec{B_0} \cdot \Vec{I_2} +\Vec{S} \cdot {A}_{^{13}C} \cdot \Vec{I_2} \label{H13C},
\end{align}
\end{subequations}
where $\gamma_{nuc}$ represents the gyromagnetic ratios ($3.077$ and $10.708$ MHz/T for $^{14}N$ and $^{13}C$, respectively), $\Vec{I_1}$ ($\Vec{I_2}$) is the $^{14}N$ ($^{13}C$) nuclear spin operator, $\Vec{A}_{nuc}$ is the hyperfine interaction ($^{14}N$: $A_{\perp} = -2.14$ MHz, $A_{\parallel} = -2.70$ MHz), and P represents the nuclear quadrupole interaction ($-5.0$ MHz).~\cite{Felton2009}
We focus our study on weakly coupled $^{13}C$ nuclear bath spins.
Using Eq.~\ref{eq:NVHamil}, we determine all eigenvalues based on the observed magnetic field.
The observed states and energies are listed in Table~\ref{tb:NVEigenVals}. 
\begin{table}
\caption{State identification and energy values determined from Eq.~\ref{eq:NVHamil} based on a magnetic field of 8.306 Tesla with an offset angle of 1.5 degrees and $A_{^{13}C} = 1$ kHz. The nuclear magnetic spin value of $^{14}N$ ($^{13}C$) is shown as $m_{I1}$ ($m_{I2}$). 
} 
\begin{ruledtabular}
\begin{tabular}{cc}
\textbf{State $\ket{m_S,m_{I1},m_{I2}}$} & \textbf{Energy [MHz]} \\ \hline
{-1, +1, +1/2} & -230023.7 \\
{-1,  0, +1/2} & -229995.3 \\
{-1, -1, +1/2} & -229976.9 \\
{-1,  +1, -1/2} & -229934.8 \\
{-1,  0, -1/2} & -229906.3 \\
{-1,  -1, -1/2} & -229887.9 \\
{0,  +1, +1/2} & -73.1 \\
{0,  0, +1/2} & -42.5 \\
{0,  -1, +1/2} & -21.9 \\
{0, +1, -1/2} & 15.9 \\
{0,  0, -1/2} & 46.4 \\
{0,  -1, -1/2} & 67.0
\end{tabular}
\label{tb:NVEigenVals}
\end{ruledtabular}
\end{table}
From Table~\ref{tb:NVEigenVals}, it is seen that the $m_S = 0$ states are not evenly spaced around zero.
This spacing is induced by partial field misalignment and nuclear quadrupole interaction that mixes the states and results in twelve non degenerate energy levels.
We next calculate allowed transitions ($\Delta m_S = 1$, $\Delta m_I = 0$) and double quantum transitions ($\Delta m_S = 1$, $\Delta m_I = 1$) involving a simultaneous electron and nuclear spin flip.
We tabulate the allowed transitions and double quantum transitions involving $^{13}C$ and $^{14}N$ spin flips in Table~\ref{tb:NVTransitions}.
\begin{table}
\caption{Simulated transition energies calculated from Table~\ref{tb:NVEigenVals}.
The states involved in the transition are listed in the left and central columns, while the calculated difference is shown in the right column.
For clarity, the transition relative to the central transition ($\nu_{sim.}-\nu_{obs.}$) was tabulated ($\nu_{obs.} = 229.9528$ GHz). 
Allowed transitions ($\Delta m_S = 1$, $\Delta m_{I} = 0$) are shown in the top panel. 
The middle and bottom panel show double quantum transitions ($\Delta m_S = 1$, $\Delta m_{I} = 1$) involving a simultaneous electron and nuclear spin flip. The middle panel shows transitions involving $^{14}N$ and the bottom panel shows transitions involving $^{13}C$.
}
\begin{ruledtabular}
\begin{tabular}{ccc}
$\bf \ket{0, m_{I1}, m_{I2}}$ & {$\bf \ket{-1, m_{I1}, m_{I2}}$} & \textbf{$\Delta$ E [MHz]} \\ \hline
{ +1, +1/2} & { +1, +1/2} & -2.1 \\
{ +1, -1/2} & {+1, -1/2} &  \\
{  0, +1/2} & { 0, +1/2} & 0.0 \\
{  0, -1/2} & { 0, -1/2} &  \\
{-1, +1/2} & {-1, +1/2} & 2.1 \\
{-1, -1/2} & {-1, -1/2} &  \\ \hline
{+1, +1/2} & { 0, +1/2} & -30.6 \\
{+1, -1/2} & { 0, -1/2} &  \\
{ 0, +1/2} & {-1, +1/2} & -18.4 \\
{ 0, -1/2} & {-1, -1/2} &  \\
{-1, +1/2} & {  0, +1/2} & 20.6 \\
{-1, -1/2} & {  0, -1/2} &  \\
{ 0, +1/2} & {+1, +1/2} & 28.4 \\
{ 0, -1/2} & {+1, -1/2} &  \\ \hline
{+1, +1/2} & {+1, -1/2} & -91.1 \\
{ 0, +1/2} & { 0, -1/2} & -88.9 \\
{-1, +1/2} & {-1, -1/2} & -86.8 \\
{+1, -1/2} & {+1, +1/2} & 86.8 \\
{ 0, -1/2} & { 0, +1/2} & 88.9 \\
{-1, -1/2} & {-1, +1/2} & 91.1
\end{tabular}
\end{ruledtabular}
\label{tb:NVTransitions}
\end{table}
As seen in Table~\ref{tb:NVTransitions}, the allowed ESR transitions are spaced by the axial hyperfine coupling to $^{14}N$, contributing to the central blind spot.
As shown in the inset of Fig. \ref{fig:C13EDNMR}(a), the signals at $-30$ and $+28$ MHz are in excellent agreement with the predicted peak positions for $^{14}N$ predicted in Table~\ref{tb:NVTransitions}.
The proximity to the central spot makes identification of the peaks at $-18$ and $20$ MHz difficult, but a  dip at $-18$ MHz is in agreement with the expected peak position.
The polarity inversion of the signals is under further investigation. The observed signals are not symmetric due to the nuclear quadrupole interaction.
The main graph of Fig. \ref{fig:C13EDNMR}(a) shows the signals at $\pm88$ MHz in excellent agreement with double quantum transitions for $^{13}C$ bath spins and are well spaced from the central blind spot. 
We next repeat the measurements on additional locations to confirm the observed signals.
We adjust the mirror to position 2, $\sim 50$ $\mu m$ from position 1, and repeat EDNMR measurements.
We also include data from a separate experimental run as position 3.
For position 3, the sample was removed from the setup and replaced, resulting in a different sample location.
Fig. \ref{fig:C13EDNMR}(a) shows $^{13}C$ EDNMR signals at $\pm 88$ MHz for all positions in excellent agreement with the simulation.

We next investigate the linewidth of the $^{13}C$ signals in more detail. 
We plot the signals related to double quantum transitions in Fig.~\ref{fig:C13EDNMR} and show the transitions as a stick spectrum.
The calculated three transition frequencies are ranged by $4.3$ MHz, which is comparable to the observed linewidth.
In general, the EDNMR linewidth is dependent on a variety of factors, including both intrinsic properties, such as spin relaxation times, and experimental parameters, such as the HTA pulse length and intensity.\cite{Cox2017}
In the present case, the observed linewidth was observed to be constant when HTA pulse lengths from $300 - 1000$ $\mu$s were used, suggesting that the linewidths are broadened by internal dynamics. 
Therefore, we focus our discussion on magnetic dipole coupling from surrounding spins which can contribute to the observed linewidth.
In general, the magnetic field at an "A" spin fluctuates due to the interaction with random spin flips of dipolar-coupled "B" spins. 
When the concentration of "B" spins is sufficiently dilute, this interaction broadens the linewidth of the "A" spin by inducing a distribution of Larmor frequencies.
In this case, the Larmor frequency fluctuations ($\Delta \omega$), at an "A" spin from the j-th dipolar coupled "B" spins may be written as:
\begin{equation}
    \Delta \omega_j = \gamma_a \delta b_j = \frac{\mu_0 \gamma_a \gamma_b \hbar}{ 4 \pi} \frac{(1-3 \cos^2\theta_j)m_j}{r^3_j},
    \label{eq:BVariance}
\end{equation}
where $\gamma_a$ ($\gamma_b$) is the gyromagnetic ratio of the "A" ("B") spin, $\mu_0$ is the permeability of free space, and $\hbar$ is the reduced planck constant.
The spin state of the j-th spin is given by $m_j$ ($m_j = \pm 1/2$ for an $S = 1/2$ spin) with $\theta_j$ representing the angle between the vector joining the spins, $r_j$, and the applied magnetic field. 
Now by considering that "B" spins are randomly distributed and the populations of the up- and down-states of "B" spins are equal, we can average $\Delta \omega_j$ by considering the probability of finding a spin at the j-th position and integrating over possible angles and spin states.~\cite{Mims1968, Klauder1962, Abragam1961}
The integral gives the full-width at the half-maximum of the Lorentzian function as a linewidth ($\Delta \omega$), which may be written as: 
\begin{equation}
   \Delta \omega = \sum_j \Delta \omega_j =  \frac{ 2 \pi  \mu_0\hbar}{ 9 \sqrt{3}}\gamma_a \gamma_b n,
   \label{eq:LinewidthC13}
\end{equation}
where n is the concentration of "B" spins in units of spins per cubic meter.
In the case of the present EDNMR study, "A" spin is the NV center and "B" spins are surrounding paramagnetic spins such as P1 centers and $^{13}C$ nuclear spins.
The concentration of nitrogen in the present sample was estimated  to be $\sim$ 70 ppm from a 230 GHz pulsed ESR measurement of the P1 center's $T_2$ ($T_2 = 1.07 \pm 0.01$ $\mu$s; data not shown).\cite{Stepanov2016}
Using 70 ppm for the concentration of P1 centers, we obtained $\Delta \omega = 3.2$ MHz.
As shown in Fig.~\ref{fig:C13EDNMR}, the simulated peaks with the three resonance frequencies and $\Delta
 \omega$
gives excellent agreement with the observed data.
Furthermore, the observed linewidth is in excellent agreement with the linewidth of the lower NV resonance (Fig.~\ref{fig:NVCharacterization} (a)) and with previous work on type-Ib diamonds.~\cite{Stepanov2016}
The use of high purity, isotopically purified diamonds with low concentrations of paramagnetic spins can be used to further improve the spectral resolution and is the subject of current work.
For example, we note that Eq.~\ref{eq:LinewidthC13} predicts $\Delta \omega \sim 0.2$ MHz from dipolar broadening due to natural abundance $^{13}C$.

We next discuss measurements on sample 2. All measurements previously discussed were repeated on sample 2. From  measurement of both the lower and upper ODMR transitions, the magnetic field was determined to be 8.306 T with a polar offset angle of $1.88 \pm 0.03^o$. Rabi oscillations showed a $\pi$ pulse length of $1.6$ $\mu$s and the $T_1$ relaxation time was measured as $3.8 \pm 0.3$ ms (data not shown).
As seen in Fig. \ref{fig:C13EDNMR}(b), EDNMR was measured at three different locations on sample 2, with EDNMR signals from $^{13}$C resolved at $\pm 88$ MHz in each location. The observed signals are in excellent agreement with the expected peak positions. The slight variation in the observed height and width from sample 1 indicates small sample to sample variation.
The inset shows EDNMR signals resolved from $^{14}$N. Clear signals are resolved at $-31$ and $+28$ MHz in excellent agreement with the simulated peak positions and sample 1.

In summary, we have demonstrated pulsed ODMR on an ensemble system of NV centers at 8.3 Tesla and 230 GHz. 
Ensemble NV centers were utilized to perform pulsed EDNMR with optical readout of the spin population. 
EDNMR signals were resolved from $^{13}C$ bath spins with the linewidth limited by the concentration of paramagnetic impurities. 
This work provides a clear demonstration of NV center detected EDNMR, and establishes groundwork for the implementation of NV-detected NMR at higher magnetic fields, with shallow NV centers, and for the study of nuclei with a variety of gyromagnetic ratios. 
EDNMR can resolve spins whose gyromagnetic ratios shift the resonance from the central blind spot. 
Nuclei with large gyromagnetic ratios, such as $^{1}H$ and $^{19}F$, are excellent candidates for future research.
Signals from bath $^{13}C$ spins were resolved in this work. 
From previous measurements, it is known that weakly coupled $^{13}C$ hyperfine interaction is on the order of $10-100$ kHz. \cite{Zopes2018,Bradley2019} 
Based on a dipolar calculation, a hyperfine coupling of more than 10 kHz is expected for surface protons within 8 nm of NVs.
With the fabrication of NVs with $T_1$ times of a few ms, stable photoluminescence, and a depth of at least 8 nm,\cite{Ishiwata2017,Osterkamp2019,Loretz2014} NV-NMR of protons at the diamond surface will be detectable with the presented EDNMR technique.
Chemical functionalization techniques can be used to bring spins of interest within close proximity of shallow NV centers.~\cite{Romanova2013, Akiel2016}
Furthermore, the described technique is limited by the comparative length of the HTA pulse relative to $T_1$ relaxation. 
As $T_1$ can be extended up to several seconds at cryogenic temperatures, this technique can utilize a long HTA pulse to perform measurements at higher fields and frequencies where microwave power is often limited.~\cite{Takahashi2008, Jarmola2012}
With the development of suitable pulsing techniques, this method will enable measurements in higher magnetic fields, such as those in the National High Magnetic Field Laboratory.~\cite{Zvyagin2004,Stoll2011}

\section{Acknowledgements}
This work was supported by the National Science Foundation (CHE-2004252 with partial co-funding from the Quantum Information Science program in the Division of Physics), the USC Anton B. Burg Foundation, and the Searle scholars program (ST). L.M.S. thanks support from USC Dornsife Chemical Biology Training Program.
Zaili Peng is acknowledged for helpful discussions regarding EDNMR.

%
%
\smallskip
\noindent{\bf DATA STATEMENT}\newline
The data that support the findings of this study are available from the corresponding author upon reasonable request.


%

\end{document}